\documentclass[conference]{IEEEtran}
\IEEEoverridecommandlockouts
\usepackage{cite}
\usepackage{amsmath,amssymb,amsfonts}
\usepackage{algorithmic}
\usepackage{graphicx}
\usepackage{textcomp}
\usepackage{xcolor}
\def\BibTeX{{\rm B\kern-.05em{\sc i\kern-.025em b}\kern-.08em
    T\kern-.1667em\lower.7ex\hbox{E}\kern-.125emX}}
\usepackage{ifthen}
\usepackage{xcolor}
\usepackage{amsmath}
\usepackage{xspace}
\usepackage{amssymb} 
\usepackage{multirow}
\usepackage{url}
\usepackage{hyperref}
\usepackage{array}
\newcolumntype{P}[1]{>{\centering\arraybackslash}p{#1}}

\usepackage{url}

\newcommand{\exclude}[1]{}
\newcommand{\showComments}{yes}
\newcommand{\note}[2]{
    \ifthenelse{\equal{\showComments}{yes}}{\textcolor{#1}{#2}}{}
}

\newcommand{\fattree}{{Clos}\xspace}

\newcommand{\para}[1]{{\textbf{{#1}}}}

\newcommand{\allreduce}{{AllReduce}\xspace}

\date{}
\usepackage{lipsum} 
\usepackage{enumitem}
\setlist{nosep} 

\begin{document}

\title{Rail-only: A Low-Cost High-Performance Network for Training LLMs with Trillion Parameters}

\author{\IEEEauthorblockN{Weiyang Wang}
\IEEEauthorblockA{MIT CSAIL \\
weiyangw@mit.edu}
\and
\IEEEauthorblockN{Manya Ghobadi}
\IEEEauthorblockA{MIT CSAIL \\
ghobadi@mit.edu}
\and
\IEEEauthorblockN{Kayvon Shakeri}
\IEEEauthorblockA{Meta \\
kvon@meta.com}
\and
\IEEEauthorblockN{Ying Zhang}
\IEEEauthorblockA{Meta \\
zhangying@meta.com}
\and
\IEEEauthorblockN{Naader Hasani}
\IEEEauthorblockA{Meta \\
naaderh@yahoo.com}
}

\maketitle

\begin{abstract}

This paper presents a low-cost network architecture for training large language models (LLMs) at hyperscale. We study the optimal parallelization strategy of LLMs and propose a novel datacenter network design tailored to LLM's unique communication pattern. We show that LLM training generates sparse communication patterns in the network and, therefore, does not require any-to-any full-bisection network to complete efficiently. As a result, our design eliminates the spine layer in traditional GPU clusters. We name this design a \textit{Rail-only} network and demonstrate that it achieves the same training performance while reducing the network cost by 38\% to 77\% and network power consumption by 37\% to 75\% compared to a conventional GPU datacenter. Our architecture also supports Mixture-of-Expert (MoE) models with all-to-all communication through forwarding, with only 8.2\% to 11.2\% completion time overhead for all-to-all traffic. We study the failure robustness of Rail-only networks and provide insights into the performance impact of different network and training parameters. \looseness=-1

\end{abstract}

\section{Introduction} \label{sec:introduction}

Large Language Models (LLMs) are among the most complex and computationally intensive Deep Neural Networks (DNNs). The GPT3 model from 2020 already requires 355 GPU-years on Nvidia's V100 GPUs~\cite{gpt3, gpt3time}, while the recent GPT4 model is estimated to have trillions of parameters and takes months to train~\cite{openai2023gpt4, gpt4params}. 
As Moore's law slows down, the growth rate of LLM size and computation requirement exceeds the advancement of accelerators, making hyper-scale GPU datacenters inevitable. Our conversations with lead machine learning architects in the industry indicate that the next-generation LLMs likely require over 30,000 GPUs of computing power to finish training within a reasonable time.

GPU manufacturers have invested in high-bandwidth platforms, such as NVLink~\cite{nvlnvs} and Infinity Fabric~\cite{mi300x}, to enable efficient multi-GPU training. These platforms provide several Tbps of bandwidth within a few GPUs but are not scalable. To connect multiple GPU platforms, state-of-the-art approaches rely on traditional lossless network solutions, such as RDMA over converged ethernet (RoCE) or Infiniband. In particular, today's GPU clusters employ an architecture called a ``Rail-optimized" network. This architecture is derived from the classical \fattree network~\cite{fattree} to provide any-to-any connectivity to \textit{all GPUs} in a training cluster.\looseness=-1  

However, scaling a \fattree network to tens of thousands of GPUs is challenging. Previous work demonstrated that large-scale lossless networks are prone to deadlocking and PFC storms~\cite{pfc_storm, rdma_azure, ib_deadlock, bfc,ddlindc}, degrading the performance. Furthermore, as the scale increases, \fattree architectures become prohibitively costly~\cite{topoopt}. For instance, today, a full-bisection Clos fabric interconnecting 30,000 GPUs with 400~Gbps network capacity costs \$200 million. At the same time, deploying such a network requires provisioning for $\sim$4.6 megawatts of peak power consumption. Consequently, datacenter providers resort to over-subscription to tame costs and energy consumption, worsening deadlocking and degraded performance problems. 

In this paper, we show that efficiently training LLMs \textit{does not} require any-to-any connectivity across all GPUs in the network, even for DNNs with sparsely gated Mixture-of-Expert (MoE) layers, which generate all-to-all communication (\S\ref{sec:moe_traffic}). As a result, we propose an immediately deployable solution to lower the cost and energy consumption of LLM datacenters with commodity electrical switches. To do so, we make two primary contributions. First, we analyze the traffic pattern of training LLMs (\S\ref{sec:traffic_ana}). We demonstrate that with an optimal parallelization strategy, an LLM training workload requires high-bandwidth any-to-any connectivity \textit{only within a small subset of GPUs}, and each subset fits within a single GPU platform, such as an Nvidia DGX server. Across the platforms, most communication occurs between \textit{a few GPU pairs with the same rank} throughout the cluster. As a result, the conventional any-to-any approach for building Clos networks adds unnecessary complexity, cost, and power consumption for distributed LLM training.

Motivated by the above observations, we then propose a low-cost network architecture that accurately reflects LLM communication requirements, called \textit{Rail-only} (\S\ref{sec:system-arch}). Instead of forming a \fattree to support any-to-any communication, as advocated by major GPU manufacturers~\cite{dgxh100}, our architecture removes the spine layer of switches and only connects sets of GPUs with significant network traffic. Hence, compared to the state-of-the-art Rail-optimized design, our network architecture removes the network equipment that does not carry significant traffic and achieves the same performance as a Rail-optimized network. We provide routing strategies that impose minimal performance overhead for all-to-all communication. We also analyze our design's fault-tolerance properties and provide recovery methods from failure cases (\S\ref{sec:fail_tol}). \looseness=-1

We evaluate the performance of our Rail-only network architecture using an analytical formulation and provide insights into the performance impact of different network and training parameters. We compare the cost and power consumption of a Rail-only network to a full-bisection bandwidth any-to-any \fattree network and show that our LLM-centric network architecture reduces the network cost and power by 38\%--76\% and 37\%--75\%, respectively (\S\ref{sec:cost}). Moreover, we show that a Rail-only network achieves the same performance as a Rail-optimized cluster for LLMs without MoE layers. Finally, we demonstrate that a Rail-only interconnect only incurs 8.2\%--11.2\% throughput overhead for LLMs with MoEs that require all-to-all traffic.  \looseness=-1

\section{Background} \label{sec:background}

\subsection{Intra-platform Connectivity: High-bandwidth Domain}
The rise of resource-intensive ML workloads led to the dominance of GPU-centric platforms optimized for multi-GPU workloads. To accommodate the communication demand, these platforms use high-bandwidth local interconnects within a local domain of GPUs. Depending on the manufacturer, these GPU-centric platforms differ in computing FLOPs, GPU and CPU architectures, or even physical interconnect technology. However, these platforms all share a unifying property: they provision  \textit{several Tbps of internal bandwidth} across GPUs. 

For instance, Nvidia's DGX H100 server~\cite{dgxh100} consists of eight H100 GPUs interconnected with NVSwitches, providing 3.6~Tbps of non-blocking bandwidth internally. The GB200 NVL72 computer announced recently connects 36 GB200 Superchips with fifth-generation NVLink within a rack at 7.2~Tbps per GPU~\cite{gb200nvl72}. The AMD MI300X platform, on the other hand, employs AMD's Infinity Fabric to connect eight MI300X accelerators in a full-mesh topology with 3.6~Tbps of bandwidth per GPU~\cite{mi300x}. Similar platforms such as Nvidia's DGX GH200 Super Computer have utilized multi-tiered NVSwich topologies to scale the platform's size up to 256 GPUs while maintaining 3.6~Tbps full-bisection intra-GPU bandwidth~\cite{gh200}.
This paper refers to a platform with Tbps internal bandwidth connectivity as a ``high-bandwidth (HB) domain", and the corresponding interconnect as HB interconnects (HBI).

\subsection{Inter-platform Connectivity: NIC Domain}

While GPU-centric platforms provide high internal bandwidth using NVLink or Infinity Fabric technologies, they can only scale to a limited number of GPUs. To expand beyond a single platform, operators rely on traditional network technologies, such as Ethernet or Infiniband, to connect the NICs of different platforms. This paper refers to the inter-platform network as the ``NIC domain." 

The state-of-the-art interconnection in the NIC domain is based on a well-known network architecture called a \textit{Rail-optimized network}~\cite{nccl212}. This architecture is ubiquitously used for high-performance computing (HPC) workloads. As we discuss next, Rail-optimized networks are better suited for DNNs than conventional CPU-centric datacenter networks. However, given that Rail-optimized networks are primarily designed for HPC workloads, they miss a significant opportunity to further leverage the unique traffic patterns of LLM training workloads (\S\ref{sec:traffic_ana}).\looseness=-1

First, let us consider a conventional datacenter design specialized to serve unpredictable and bursty CPU-heavy workloads. This architecture, known as a \fattree network~\cite{fattree, fbdcarch}, provides any-to-any connectivity between server pairs. \fattree networks are well-studied in the system and networking community and are the de facto infrastructure for storage, cloud, and map-reduce workloads.  \looseness=-1
 
The Rail-optimized network for GPU training clusters evolves from the datacenter \fattree network~\cite{nccl212, dgxh100archdoc}, illustrated in Figure~\ref{fig:Rail-optimized}. For a GPU platform with an HB domain of size $K$, there are $K$ total rails, where a \textit{rail} comprises GPUs with the same local rank that belong to different HB domains~\cite{multirail}. A Rail-optimized network places these GPUs under the same set of switches, which we denote as rail switches. Figure~\ref{fig:Rail-optimized} highlights rail one and rail $K$ in red and yellow colors, respectively. Connecting same-rank GPUs to the same rail switches ensures the lowest possible latency across them. Such connectivity is desirable because an optimal DNN parallelization strategy concentrates its NIC domain traffic between GPUs with the same local rank~\cite{nccl212}.

Rail-optimized architectures enjoy low latency between GPUs in the same rail. The rest of the network employs layers of spine switches to connect the rail switches to form a full-bisection any-to-any \fattree network topology. This network ensures that any pair of GPUs in different HB domains can still communicate at the network line rate of hundreds of Gbps. For instance, traffic between \texttt{GPU 1, Domain 1} and \texttt{GPU 1, Domain 2} traverses through Rail Switch 1 only, while traffic between \texttt{GPU 1, Domain 1} and \texttt{GPU 2, Domain 2} goes through the respective rails and the spine switches. \looseness=-1

While the Rail-optimized network architecture takes advantage of the strong locality of DNN training traffic by connecting the same-rank GPUs with the same ToR switch, it overlooks a fundamental question: Are the spine switches necessary? In the next section, we analyze LLM training traffic in greater detail to explore the potential for a spineless network architecture design.

\section{LLM Traffic Pattern Analysis}
\label{sec:traffic_ana}

\subsection{Traffic Pattern of MegatronLM}
We now analyze the traffic pattern generated by LLMs with hybrid data, tensor, and pipeline parallelism by computing the network transfer sizes from the model hyperparameters and the parallelization strategy. We first look at a series of GPT models with 145.6 billion, 310.1 billion, 539.6 billion, and 1 trillion parameters described in Table~1 of MegatronLM~\cite{narayanan2021efficient} paper, distributed across up to 3072 GPUs. The models are distributed in a cluster of up to 384 DGX A100 servers with an HB domain of size eight. Our analysis uses the same parallelization strategy from MegatronLM to ensure optimal GPU utilization. We use the ring-based collective communication since it is bandwidth-optimal and the default algorithm in NCCL. 

\begin{figure}[t]
\centering
\includegraphics[width=\columnwidth]{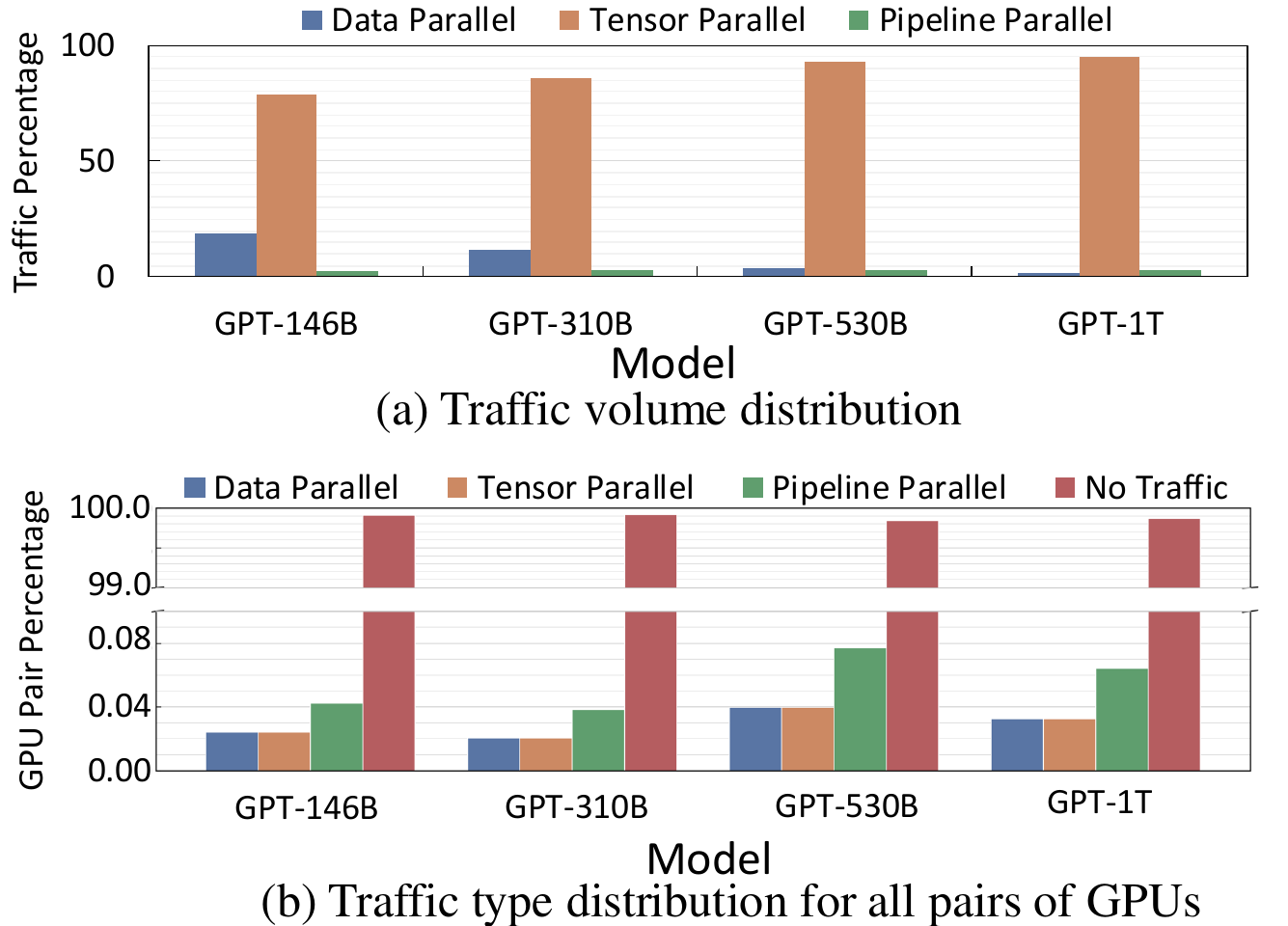} 
\caption{(a) The traffic volume from different parallelization dimensions; (b) The communication type across all GPU pairs.}
\label{fig:traffic_dist}
\end{figure}

There are three primary types of communication: AllGather and ReduceScatter traffic from tensor parallelism (TP), \allreduce traffic from data parallelism (DP), and point-to-point traffic from pipeline parallelism (PP). Figure~\ref{fig:traffic_dist}a illustrates the volume percentage for each type of communication for one training iteration. Figure~\ref{fig:traffic_dist}b shows the communication type distribution across all GPU pairs. 

The TP traffic happens within GPUs participating in a TP rank, which occupies an HB domain. The DP and PP traffic happen in the NIC domain, and their volume is significantly smaller than TP traffic, as illustrated by Figure~\ref{fig:traffic_dist}a. While traffic from different parallelism does not overlap between different pairs of GPUs, Figure~\ref{fig:traffic_dist}b indicates that over 99\% of GPU pairs carry \textit{no traffic} and less than 0.04\% of GPU pairs carry TP traffic. Simultaneously, Figure~\ref{fig:traffic_dist}a suggests TP traffic accounts for over 75\% of the total transmitted data. Recall that TP traffic stays within HB domains, suggesting efficient usage of HB domain bandwidth and low demand in the NIC domain. This pattern is consistent across all GPT models we studied, indicating that building a GPU datacenter with any-to-any connectivity on top of HB domains for LLM models is excessive. 
\begin{figure}[t]
\centering
\includegraphics[width=\columnwidth]{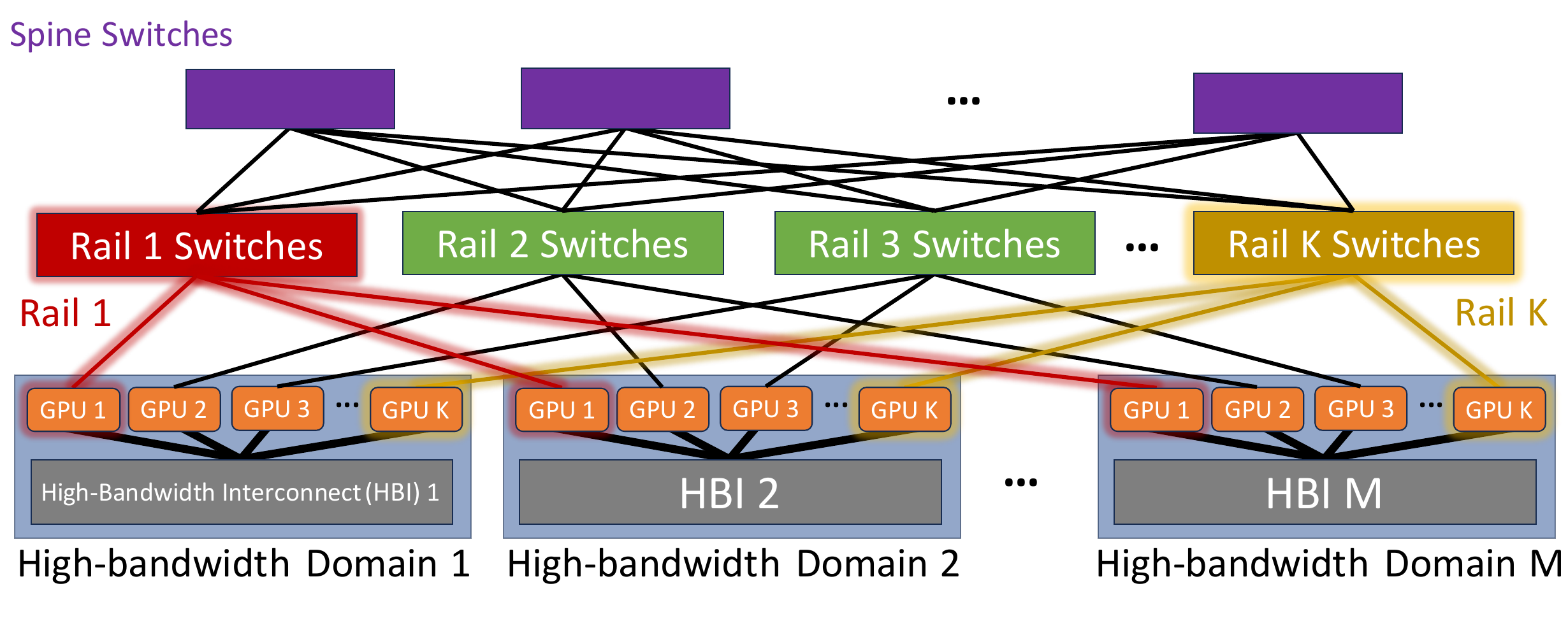}
\caption{A GPU datacenter with Rail-optimized, any-to-any \fattree networks~\protect\cite{dgxh100archdoc}.}
\label{fig:Rail-optimized}
\end{figure}

\subsection{Traffic in the NIC Domain for LLMs}

\begin{figure}[t]
    \includegraphics[width=\columnwidth]{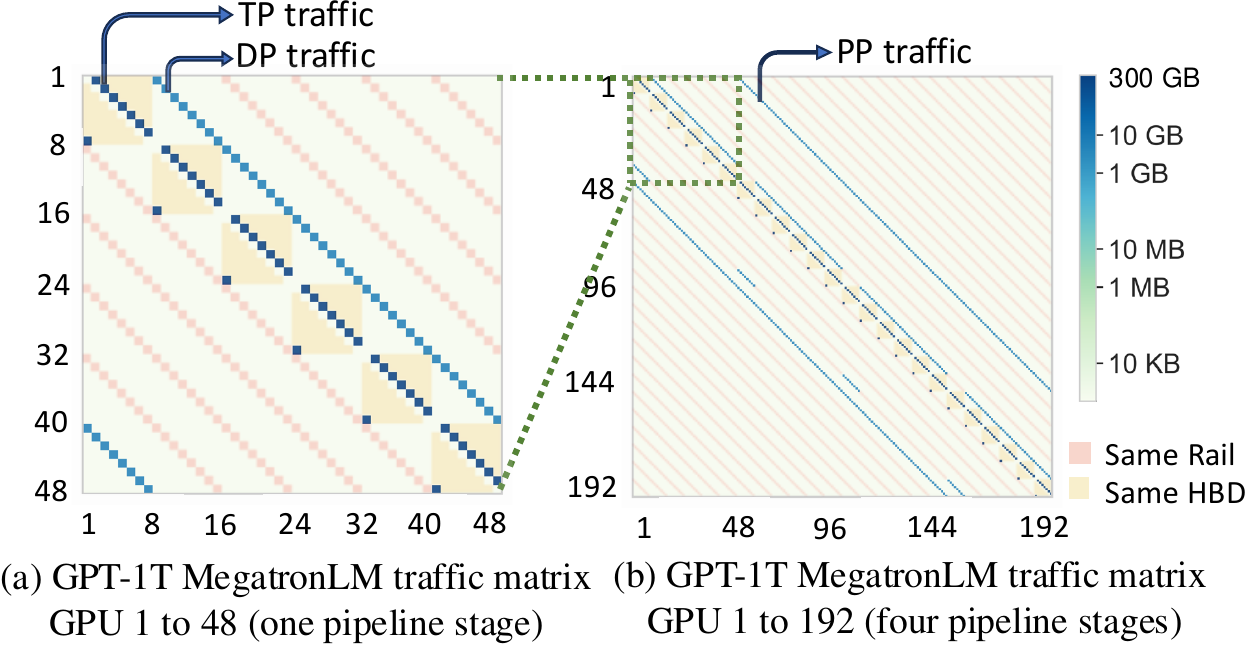}
    \caption{Traffic heatmaps for GPT-1T in MegatronLM~\cite{narayanan2021efficient}. Highlights show GPUs in the same HB domains and rails.}
    \label{fig:heatmap}
\end{figure}

The parallelization strategy employed in MegatronLM induces an insignificant amount of traffic in the NIC domain compared to the HB domains. Figure~\ref{fig:heatmap} shows the traffic heatmap for training the GPT-1T model. In this plot, every consecutive set of eight GPUs resides within the same HB domain (highlighted in orange), and GPUs with a distance of 8 between them belong to the same rail (highlighted in pink). Figure~\ref{fig:heatmap}a demonstrates the traffic pattern within one pipeline stage, while Figure~\ref{fig:heatmap}b shows the traffic across the first four pipeline stages. The traffic volume is significant ($\sim$300~GB across GPU pairs) in an HB domain, while the communication drops to only about 6~GB in the NIC domain. Furthermore, the transfers in the NIC domain never traverse through the spine switches -- these network transfers only happen within a rail. \looseness=-1

We argue that \textit{all} LLMs without sparse MoE layers distributed with an optimal parallelization strategy always induce sparse, low-volume traffic in \textit{within the rails}. By design, the only traffic exiting the HB domains is point-to-point traffic from pipeline parallelism or collective communication traffic (AllGather, ReduceScatter, and AllReduce) from TP and DP. 

\begin{figure*}[t]
\centering
\begin{minipage}{.95\textwidth}
\includegraphics[width=\textwidth]{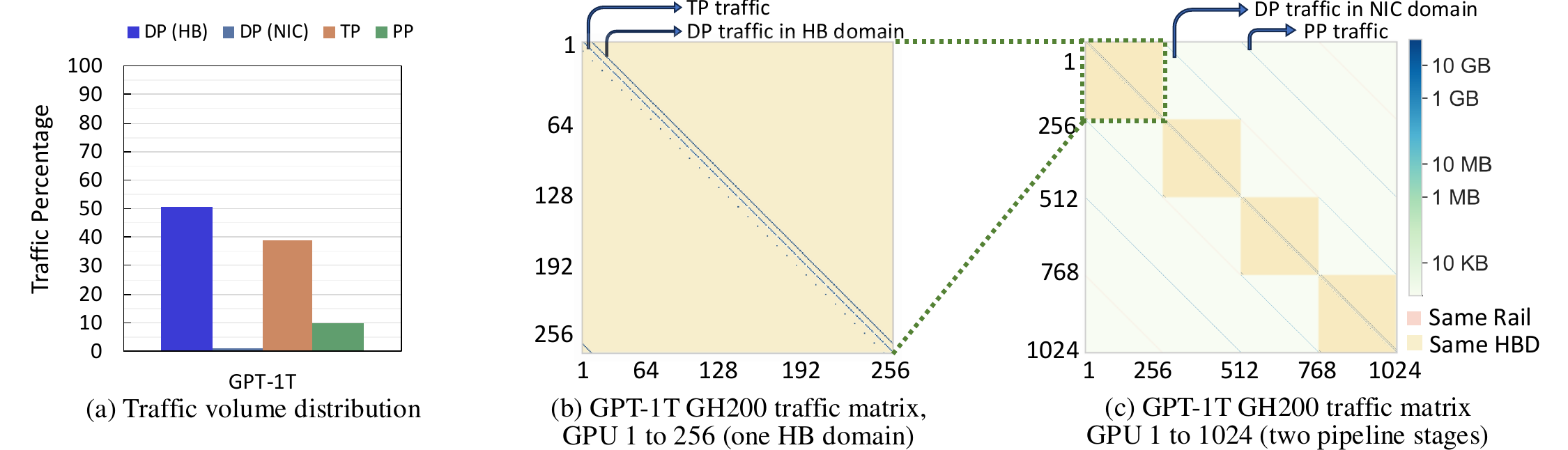}
\caption{Traffic distribution and heatmaps for GPT-1T, distributed on 16 DGX GH200s. Note that DP (NIC) accounts for 0.8\% of the total traffic percentage. The "Same-Rail" legend on Figure~\ref{fig:traffic_dist_gh200} appears for GPUs whose ranks are 256 apart.}
\label{fig:traffic_dist_gh200}
\end{minipage}
\end{figure*}

For PP, due to the symmetry of LLM parallelization, each pipeline stage contains the same number of GPUs. As a result, the pipeline stages can always be placed such that traffic across stages traverses the GPUs of the same rank in the NIC domain only, hence staying within the same rail. 

TP and DP can induce collective communication traffic in both the HB and the NIC domains. The examples from MegatronLM always have TP and DP contained within HB and NIC domains only, respectively. While such a partition is common for LLMs, it is not ubiquitous. For example, training a smaller model using only DP causes all GPUs to participate in the same DP rank and, thus, the same AllReduce operation across both HB and NIC domains. In these cases, the training cluster could use \textit{hierarchical collective communication} algorithms that achieve near-optimal performance. Below, we introduce these algorithms, analyze their performance, and illustrate that their traffic in the NIC domain stays within rails.

Hierarchical collective communication algorithms are designed for a multi-tiered network topology. We introduce the hierarchical AllGather algorithm here and note that for the other collectives happening in LLM training, ReduceScatter achieves the same performance by inverting the schedule of AllGather, and AllReduce is equivalent to a ReduceScatter followed by an AllGather. We focus on the bandwidth performance and ignore the latency, as the data transmission is significant during LLM training; thus, the communication runtime is bandwidth-dominated. 
Table~\ref{tab:var_anal} lists the variables used in this section. We assume the GPUs conducting an AllGather operation are arranged into an $x\times y$ grid, where each $x$ GPU belongs to the same HB domain and across $y$ total HB domains. The basic hierarchical AllGather executes in two phases: first, the algorithm collects partial data for each rail of GPUs without transferring data in the HB domain. If the total data to run AllGather is of size $D$, then the amount of data exchanged in the network by all GPUs is ${D(y-1)}/{x}$. This operation effectively creates larger data shards for the GPUs to rerun AllGather within each HB domain. Therefore, each HB domain conducts an AllGather in the second phase, inducing a total transfer of $D(x-1)$. Assume the $x$ GPUs within an HB domain have bandwidth capacity $C_F$ and $y$ GPUs in the NIC domain have bandwidth $C_S$, then the total AllGather runtime is \looseness=-1
\begin{equation}\label{eq:agtime}
    \mathtt{AGtime}(D, x, y, C_F, C_S)=\frac{(y-1)D}{xyC_S}+\frac{(x-1)D}{xC_F}
\end{equation} 
Like PP communication, by appropriately mapping the logical $x\times y$ GPUs to the GPU cluster, this algorithm only induces traffic for GPUs within the same rail. 

Figure~\ref{fig:traffic_dist_gh200}a shows the traffic pattern of training GPT-1T with hierarchical collective communication spanning both NIC and HB domains, differing from the MegatronLM cases. We compute and analyze the traffic pattern of training the GPT-1T model, with a batch size of 4096, distributed in a cluster composed of 16 Nvidia DGX GH200 supercomputers~\cite{gh200} (4096 GPUs). Each DGX GH200 supercomputer comprises a two-tier NVSwitch architecture, facilitating 3.6~Tbps GPU-to-GPU bandwidth across 256 H100 GPUs. Additionally, each GPU has a Connect-X7 HCA Infiniband network interface~\cite{gh200}, which provides 400~Gbps network bandwidth in/out of each GPU. In this setup, each DGX GH200 supercomputer forms an HB domain. Figure~\ref{fig:traffic_dist_gh200} illustrates the traffic volume percentage and heatmap in this setting. The parallelization strategy has a total data parallel degree of 64, which spans 32 GPUs in each HB domain and two HB domains across the network. Figures~\ref{fig:traffic_dist_gh200}b and~\ref{fig:traffic_dist_gh200}c illustrate the traffic heatmap of the hierarchical AllReduce algorithm, which splits the AllReduce traffic among each DP group. Note that the network traffic stays within a rail (GPUs with a distance of 256 apart). The hierarchical algorithm efficiently utilized the bandwidth in the HB domain to carry $98\%$ of the AllReduce traffic, while the network domain carries the other $2\%$.

\subsection{All-to-All Traffic for Mixture-of-Expert Models}
\label{sec:moe_traffic}
LLMs with sparsely gated Mixture-of-Expert (MoE) layers exhibit a different traffic pattern than the models described above. MoE layers provide an alternative way to increase the size of LLMs without introducing significant additional computational requirements. With MoEs, part of the model is replaced by a set of ``expert" neural networks, where a \textit{gating network} routes each token to different experts, thereby only activating part of the model. The typical parallelization strategy for MoEs is expert parallelism, where each expert is distributed to a different GPU in the cluster.

\begin{figure}[t]
\centering
\includegraphics[width=\columnwidth]{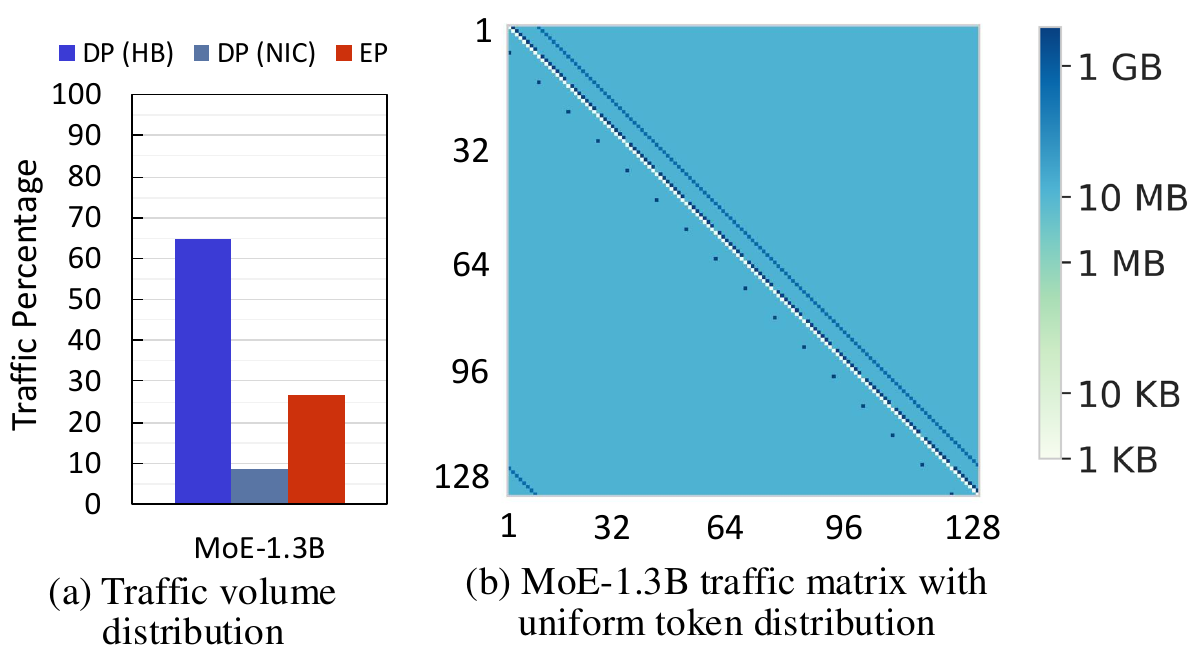}
\caption{Traffic volume distribution and heatmap for the MoE-1.3B model in DeepSpeedMoE~\cite{rajbhandari2022deepspeedmoe}, assuming uniform token distribution.}
\label{fig:moe_traffic}
\end{figure}

Unlike traditional LLMs, MoEs with expert parallelism require each expert to communicate with the rest of the model, creating a dense communication pattern. The exact traffic heatmap depends on the gating network and the token distribution. In this section, we assume a uniform token distribution for simplicity. Figure~\ref{fig:moe_traffic} shows the traffic matrix of training the MoE-1.3B model from DeepSpeedMoE~\cite{rajbhandari2022deepspeedmoe}, with 16 DGX A100 servers. The model contains 128 experts. The static part of the model uses DP, while the MoE part uses expert parallelism. Since each GPU contains a different expert, a uniform token distribution generates a uniform all-to-all traffic pattern across the network. At first glance, such traffic patterns make the spine switch in the rail-optimized network crucial, as the traffic across GPUs on different rails will traverse through spine switches. However, as we show in the next section, we do not have to rely on the spine switches: using the HB domain as a forwarding step achieves near-optimal performance.

\begin{table}[t]
\scriptsize
\centering
\caption{Variables used in Section~\ref{sec:traffic_ana} and ~\ref{sec:routing}.}
\renewcommand{\arraystretch}{1} 
\linespread{1.1}\selectfont\centering
    \begin{tabular}{|P{15mm}|P{65mm}|}
    \hline
    $x$ & GPU grid dimension in HB domains. \\ \hline
    $y$ & GPU grid dimension in NIC domains. \\ \hline
    $C_F$ & Bandwidth of HB domains. \\ \hline
    $C_F$ & Bandwidth of NIC domains. \\ \hline
    $D$ & Data transfer size between a pair of GPUs. \\ \hline 
    $T_{a2a}^{Rail-opt}$ & All-to-all traffic completion time for Rail-optimized networks. \\ [0.3ex] \hline 
    $T_{a2a}^{Rail-only}$ & All-to-all traffic completion time for Rail-only networks. \\ [0.3ex] \hline 
    \end{tabular}
    \label{tab:var_anal}
\end{table}

\section{Rail-only Network Design}\label{sec:system-arch}
Based on the observations above, this section proposes \textit{Rail-only}, a novel network architecture that diverts from the any-to-any GPU connectivity paradigm. We first introduce the architecture design of a Rail-only network. We then discuss routing policy and fault-tolerance properties of Rail-only interconnects.

\subsection{Architecture Design}
\begin{figure}[t]
\centering
\includegraphics[width=\columnwidth]{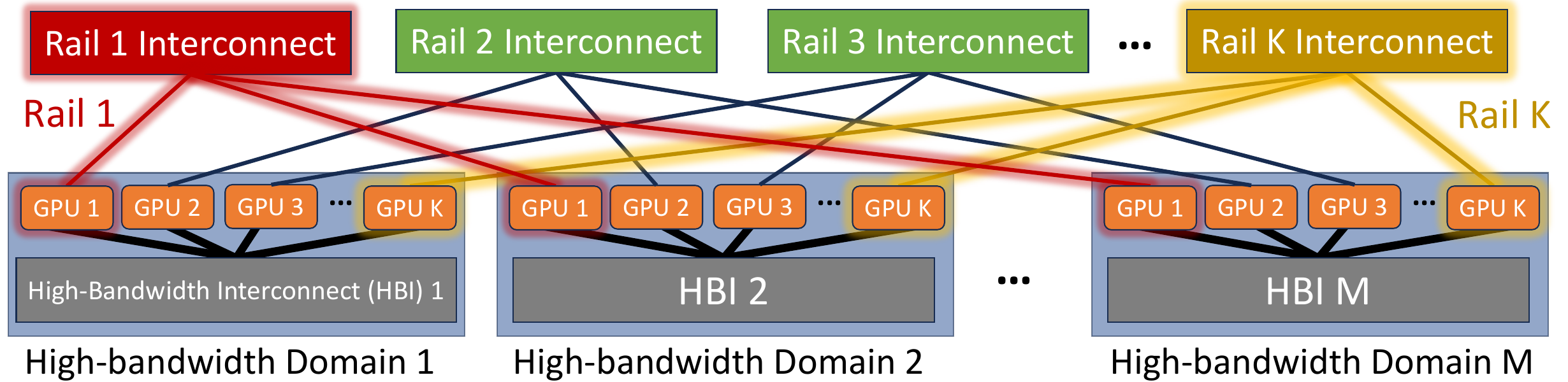}
\caption{Our proposal: a \textit{Rail-only} network providing full-bisection connectivity to individual rails.}
\label{fig:Rail-only}
\end{figure}

Figure~\ref{fig:Rail-only} illustrates our \textit{Rail-only} network architecture. Compared to a conventional Rail-optimized GPU cluster, shown in Figure~\ref{fig:Rail-optimized}, our Rail-only network keeps the HB domains but omits the full-bisection connectivity \textit{for all GPUs} in the NIC domain. Instead, we only ensure that \textit{GPUs within each rail} are connected with a full-bisection network.\looseness=-1 

A straightforward illustration of our proposed architecture is to remove the spine switches (Figure~\ref{fig:Rail-optimized}) and re-purpose the uplinks connecting rail switches to the spine as downlinks to GPUs. Hence, a dedicated \fattree network connects each rail. Compared to the Rail-optimized architecture, the Rail-only design saves the number of switches and transceivers by building multiple smaller \fattree networks for individual rails, requiring fewer layers of switches in the network.

\subsection{Routing in Rail-only Networks} \label{sec:routing}
Our Rail-only network architecture removes network connectivity across GPUs with different ranks in different rails. Such communication is still possible by forwarding the data through HB domains. For instance, a message from \texttt{GPU 1, Domain 1} to \texttt{GPU 2, Domain 2} in Figure~\ref{fig:Rail-only} can first route through the first HB domain to \texttt{GPU 2, Domain 1} and then be transmitted to the final destination through the network. Previous work has shown that such a routing scheme induces a \textit{bandwdith-tax}~\cite{rotornet, opera, topoopt}, where the physical traffic increases in the network due to forwarding. However, in this section, we show that due to the bandwidth asymmetry between the HB and the NIC domain, the performance degradation caused by bandwidth tax is negligible.  \looseness=-1

We use LLMs with MoE layers described in Section~\ref{sec:moe_traffic} as an example, as it generates a challenging all-to-all communication pattern. At first glance, this traffic pattern is challenging for the Rail-only network. Most of the all-to-all traffic requires forwarding. However, since the HB domain is much faster than the NIC domain, forwarding network traffic on the HB domain incurs a slight overhead. Below, we derive the slow-down factor for uniform all-to-all traffic using a two-step forwarding routing algorithm for the Rail-only network. Note that this strategy has already been implemented in NCCL as ``PCIe $\times$ NVLink" (PXN) to avoid congestion in cases where the spine layer of the Rail-optimized network is oversubscribed~\cite{nccl212}.\looseness=-1

We use the same variables defined in Table~\ref{tab:var_anal} in the following derivation. Consider a grid of $x\times y$ GPUs where $x$ GPUs are placed in an HB domain, and a Rail-only or Rail-optimized network connects $y$ HB domains. Recall that the HB domain has bandwidth $C_F$ while the NIC domain has bandwidth $C_S$ per GPU pair. For a Rail-optimized network, every GPU sends traffic directly to its destinations through the HB and full-bisection NIC domains. Assuming each pair of GPU communicates traffic of size $D$, the total all-to-all completion time is:
\begin{equation}
T_{a2a}^{Rail-opt}=\max(\frac{(x-1)D}{C_F}, \frac{x(y-1)D}{C_S})= \frac{x(y-1)D}{C_S}
\end{equation}
For the Rail-only network, the two-step algorithm first runs an all-to-all \textit{within each rail}, preparing each GPU to have all data to send on its rail. Then, within each rail, each GPU runs a second all-to-all \textit{in the HB domain} to finish the transfers with an effective shard size of $xD$. Note that the second step here contains the \textit{bandwidth-tax}. The total transmission time is
\begin{equation}
T_{a2a}^{Rail-only}=\frac{y(x-1)D}{C_F} + \frac{x(y-1)D}{C_S}
\end{equation}
The two terms differ by ${y(x-1)D}/{C_F}$, which is the cost of forwarding within HB domains. When $ y(x-1)\approx x(y-1)$, the slow-down factor is approximately $C_S/C_F$, which equals to $8.2\%$ and $11.2\%$ for the DGX A100 and DGX H100 generations of GPU platforms, respectively. This factor is low because of the bandwidth asymmetry between the two domains. Furthermore, this slow-down only applies to the all-to-all communication, which accounts for $27\%$ of the total traffic as shown in Figure~\ref{fig:moe_traffic}. Therefore, this small overhead is negligible by Amdah's law. We note that such properties also make our network design suitable for other classes of DNN models.   \looseness=-1

\subsection{Fault Tolerance Properties of Rail-only Networks} \label{sec:fail_tol}
Fault tolerance is crucial for large GPU clusters with long-lasting LLM training jobs. This section investigates the fault tolerance properties of  Rail-only networks compared to traditional Rail-optimized networks.

\para{Link and switch failures.} Suppose a rail switch or a link fails. GPUs connected to the failed switch or link become unavailable for both Rail-optimized and Rail-only network architectures, rendering the two designs identical regarding fault tolerance on switches and links. However, our design requires fewer switches, which naturally reduces the points of failure. Datacenter operators can add redundant capacity by including extra rail switches, and our design remains more cost-effective than the any-to-any network design. \looseness=-1 

\para{GPU platform failure.} For a GPU cluster composed of DGX-like servers, each server is an HB domain. When a server fails, the network operator migrates the task to another healthy server. The Rail-only connectivity remains the same for the new server. For a GB200-like cluster, a super-chip module is analogous to a server; thus, the failure mode is the same as a single GPU failure, which we will discuss next.  \looseness=-1

\para{Single GPU failures with idle GPU in the HB domain.} We discuss two distinct scenarios separately for single GPU failures. The first case is when another idle GPU presents the same HB domain as the failed one. In this case, a Rail-optimized network directly replaces the failed GPU with the idle one without breaking the HB domain integrity. One possible solution is to leverage optical reconfigurable switches for the Rail-only network. To improve robustness, we add a small number of optical switches between the GPU and rail switches to allow the dynamic reconfiguration of rails. When a GPU fails, the optical switch reconfigures to bring a healthy, idle GPU to replace the failed one. This approach is conceptually similar to the failure recovery mechanism of network designs that primarily uses optical-reconfigurable switches~\cite{jouppi2023tpu, topoopt, jupiterevolving}. We leave an in-depth analysis of rail-only with optical switch to future work.

\para{Single GPU failure in fully occupied HB domains.} Another failure mode occurs when a GPU fails in a fully occupied HB domain and requires a substitute GPU from different HB domains. In this case, the Rail-only design prevents migrating the task on the failed GPU to another idle one in the cluster, which is possible in a Rail-optimized network. However, such a solution is undesirable even with the Rail-optimized network. The new GPU no longer belongs to the same HB domain as the failed one, creating a bottleneck that slows the HB domain into a NIC domain. Instead, we propose two solutions. For platforms with small HB domains, we migrate the tasks on the entire HB domain with the failed GPU to a new one. For larger HB domains (e.g., DGX GH200 supercomputers with $K=256$), these HB domains comprise a multi-tiered topology with an optical core-layer~\cite{gh200}. One potential approach is to add optical switches, like in the previous failure case. When a GPU failure occurs, the optical switch reconfigures, replacing a small set of GPUs (including the failed one) with healthy ones, thus maintaining the integrity of an HB domain.  \looseness=-1

\section{Evaluation} \label{sec:evaluation}

\subsection{Iteration Time Modeling} \label{sec:accuracy_model}
We evaluate our Rail-only network design's performance through an analytical model of the training iteration time. Our analytical model considers the critical path for LLM training with TP, DP, and PP, similar to the approach in Calculon~\cite{calculon}. 

We demonstrate the accuracy of our analytical model by comparing its estimation of hardware FLOPs utilization (HFU) to that of published results in the literature. The HFU refers to the hardware's floating point operation performed in an iteration over the peak floating point operations. Prior work provided the full set of hyperparameters in their evaluation setup, enabling us to compare the estimated HFUs from our analytical model to the ground truth~\cite{korthikanti2022reducing}. In our evaluations, we assume a cluster of 1 to 280 DGX A100 servers. 
To compute the total required FLOPs for training per iteration of a DNN model, we use the same formula proposed by Korthikanti et al.~\cite{korthikanti2022reducing}.

\begin{figure}[t]
\centering
\includegraphics[width=\columnwidth]{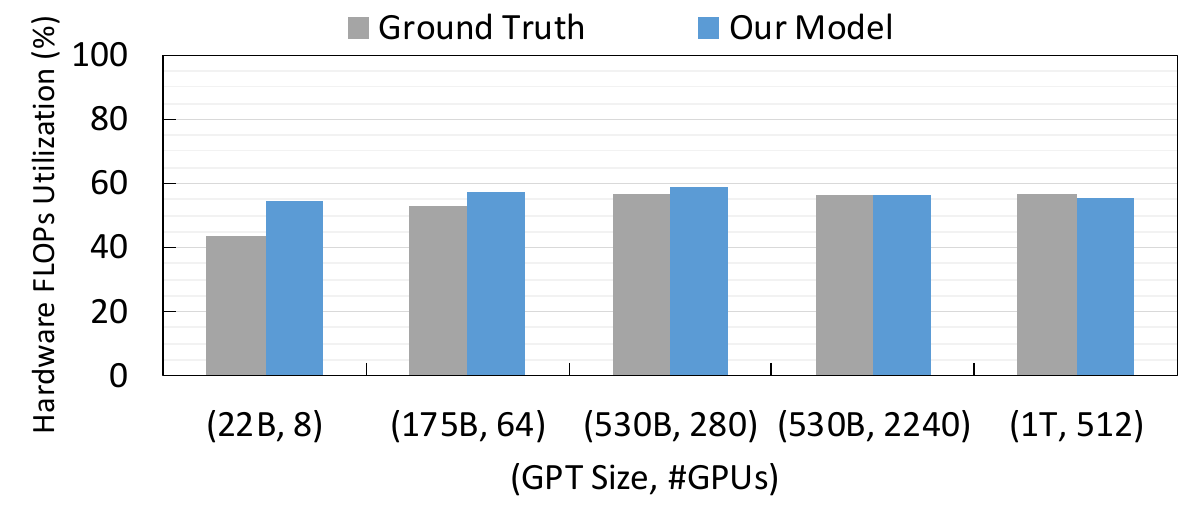}
\caption{HFU comparison between the ground truth~\cite{korthikanti2022reducing} and our formulation.}
\label{fig:hfu}
\end{figure}

Figure~\ref{fig:hfu} compares the HFUs approximated by our analytical model with the ground truth for different GPT models and cluster scales. For GPT-1T, our computed HFU only differs from the ground truth by $1.8\%$. The discrepancy between our analytical model and the ground truth comes from our idealistic modeling of GPU computation and communication, assumptions on how computation and communication overlap, and underestimation of memory overhead. Such discrepancy goes up as the model size decreases. For the GPT-22B model, our estimation error is $10.8\%$ compared to ground truth MFU. 
The rest of this section utilizes our analytical model to estimate the training iteration time of Rail-only interconnects.

\subsection{What is the Optimal Size of an HB domain?} \label{subsec:hbd-size}

\begin{figure}[t]
\centering
\includegraphics[width=\columnwidth]{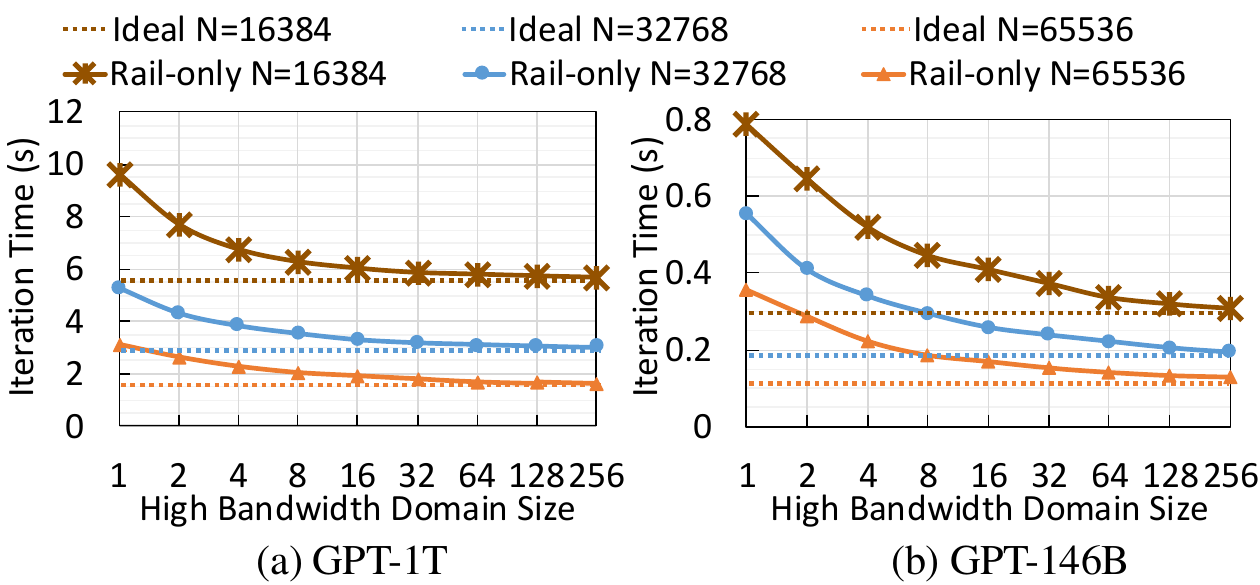}
\caption{Iteration time as HB domain size changes.}
\label{fig:iter_time_dom_sz}
\end{figure}

Intuitively, increasing HB domain size reduces the inter-platform network overhead during training. This section answers the following question: \textit{what is the optimal size of an HB domain in a Rail-only interconnect?} In Figure~\ref{fig:iter_time_dom_sz}, we vary the HB domain size ($K$) and plot the training iteration time for GPT-1T and GPT-146B MegatronLMs for clusters with 16384, 32768, and 65536 H100 GPUs. The global batch size in this evaluation is 4096 and 1024 for GPT-1T and GPT-146B, respectively. We enumerate all possible parallelization strategies for each cluster size and use the optimal parallelization strategy found in our analytical model, using the bandwidth and computational ability parameters of DGX GH200. In addition, to capture the ideal iteration time, we also compute training iteration time where a full-bisection monolithic NVLink fabric connects every GPU (i.e., the case where $K=N$, where $N$ is the total GPU count). \looseness=-1

As depicted in Figure~\ref{fig:iter_time_dom_sz},  
iteration times decrease as the HB domain size increases, indicating that larger HB domains reduce the network overhead during training. In both GPT models, the iteration time achieved with an HB domain size of 256 is nearly optimal. Compared to the ideal case (all GPUs are under a monolithic HB domain), GPT-146B with an HB domain of 256 is $4.1\%$ slower, while GPT-1T is $0.9\%$ slower. However, the \textit{marginal gain} decreases as the HB domain size increases.
For the larger GPT-1T model, the train iteration time plateaus above 32 GPUs due to Amdhahl's law, suggesting diminishing returns from the increased cost of bigger HB domains. 

For a smaller GPT-146B model, shown in Figure~\ref{fig:iter_time_dom_sz}b, the marginal gain of increasing HB domain size is higher than that of GPT-1T. Providing an HB domain of size eight reduces the iteration time by $43.3\%$ compared to the HB domain of size one, while increasing the HB domain size from 8 to 256 further achieves a $30.6\%$ reduction in iteration time. The more significant marginal gain for smaller LLMs incurs more communication overhead when distributed to the same cluster than larger models. This effect arises from how computation and communication costs scale as LLM grows. The communication requirement increases linearly with the model's hidden size and sequence length. On the other hand, the model FLOPs increase quadratically with these two parameters, as indicated by previous work~\cite{narayanan2021efficient}. Therefore, we conclude that the optimal HB domain size depends on the size of the GPT model.

\subsection{Impact of HB domain and Network Bandwidth}
\begin{figure}[t]
\centering
\includegraphics[width=\columnwidth]{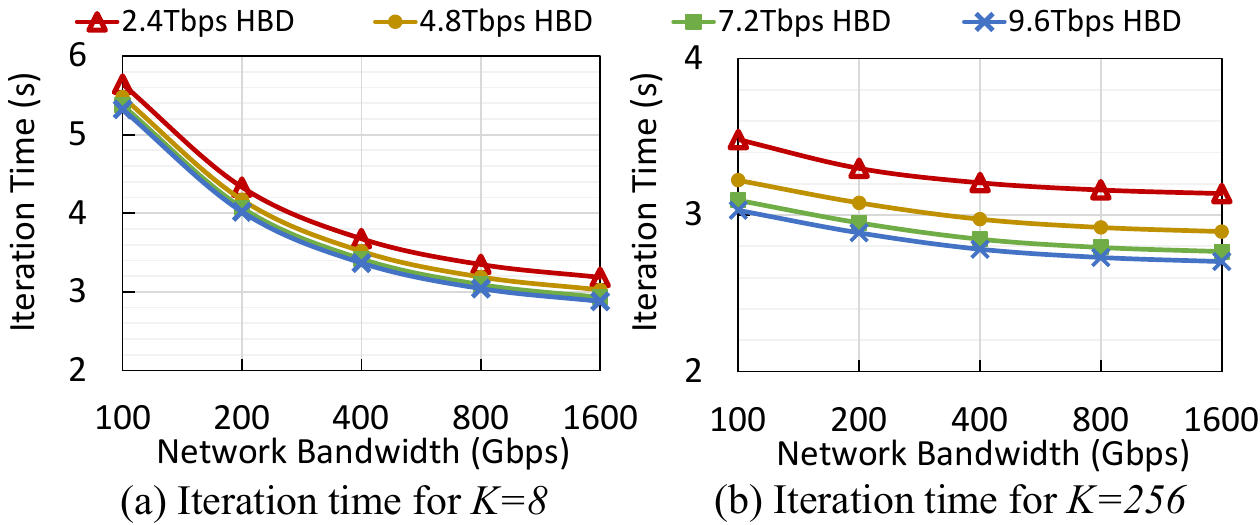}
\caption{Iteration time of GPT-1T as HB domain bandwidth and network bandwidth varies for different HB domain sizes.}
\label{fig:iter_time_bw}
\end{figure} 

The available bandwidth of HB and NIC domains determines the communication time during training. We analyze the impact of these bandwidths on the iteration time of a GPT model with one trillion parameters. Figure~\ref{fig:iter_time_bw}a and~\ref{fig:iter_time_bw}b show the iteration time variation for different HB domain bandwidths (different lines) and network bandwidths in the rails (on the $x$-axis) for HB domain size $K=8$ and 256, respectively. As expected, the iteration time decreases when either bandwidth increases. However, the $K=8$ case is less sensitive to the HB domain bandwidth. Increasing per-GPU bandwidth by a factor of four (from 2.4~Tbps to 9.6~Tbps) only improves the iteration time by $8.0\%$ on average for $K=8$, compared to the improvement of $13.3\%$ for $K=256$. On the other hand, larger HB domain sizes are more pronounced on network bandwidth improvement. Increasing the bandwidth from 100~Gbps to 400~Gbps, results in $35.9\%$ improvement for $K=8$ but only $8.0\%$ for $K=256$. Hence, improving the HB domain bandwidth is more beneficial than the network bandwidth for LLMs as future HB domain size increases. \looseness=-1

\begin{figure}[t]
\centering
\includegraphics[width=0.96\columnwidth]{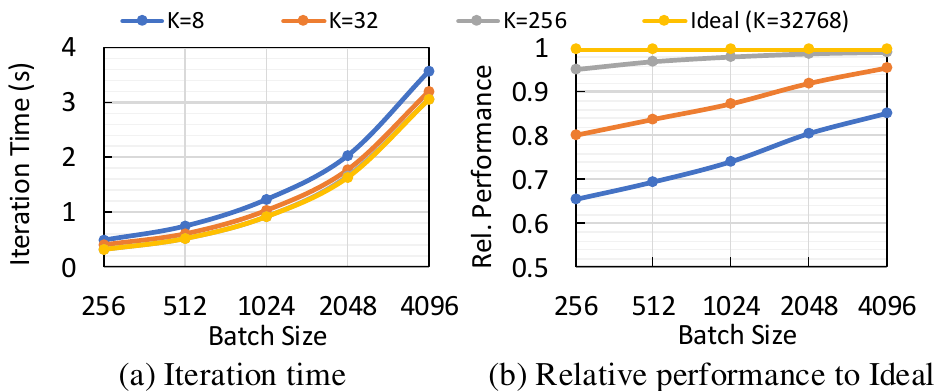}
\caption{Iteration time and relative performance to the ideal case, as batch size changes, for GPT-1T.}
\label{fig:iter_time_palm}
\end{figure}

\subsection{Impact of Batch Size on Network Design} \label{subsec:batch-size}

While the batch size is typically an ML-centric metric for optimization, our analysis indicates that the impact of batch size on the total training time goes beyond the total number of iterations required for convergence. To further understand such impact, we analyze the iteration time of a GPT-1T model on a 32768 GPU cluster while changing the HB domain size from $K=8$ to $32768$. In this study, we vary the global batch size from 256 to 4096. Figure~\ref{fig:iter_time_palm}a plots the change in iteration time as the batch size varies. The iteration time exhibits a similar trajectory for all HB domain sizes; however, the \textit{relative performance} (the ratio of the iteration time for an HB domain size to that of the ideal case) improves as the batch size increases. Figure~\ref{fig:iter_time_palm}b represents this trend. When $K=256$, the relative performance increases from $95\%$ to $99\%$ as the batch size increases from 256 to 4096 sequences. The iteration time advantage is prominent when the HB domain size is small. For $K=8$, increasing the batch size from 256 to 4096 improves the relative performance from $65\%$ to $85\%$, suggesting a larger batch size is preferable for a cluster with a smaller HB domain. Prior studies have shown that LLM training benefits from a larger batch size~\cite{kaplan2020scaling, gpt3}, especially for bigger models, making it a perfect fit for our Rail-only design. \looseness=-1

\subsection{Network Cost and Power Analysis} \label{sec:cost}

Our Rail-only network architecture judiciously reduces the network resources for LLM training by eliminating unused network hardware. This section compares our proposed approach's network cost and power with the state-of-the-art Rail-optimized GPU clusters. We calculate the number of switches ($N_{SW}$) and transceivers ($N_{TR}$) required for each network design and derive the network equipment cost and peak power consumption based on numbers reported in prior work and by vendors~\cite{topoopt, nvidia400gpower, qm7900spec}.\footnote{Cost: $Price_{TR}=\$199$ per transceiver, $Price_{SW}=\$694$ per switch port for 400~Gbps; Power: $Power_{TR}=9$W per transceiver, $Power_{SW}=18$W per switch port}
We enumerate the number of switches and transceivers required to build the state-of-the-art network architecture and our proposed architecture in Table~\ref{tab:nw_swich}, accounting for variable cluster sizes and network switch radix. 
For each architecture, we build a minimum layer \fattree network with the provided switch radix and calculate the minimum number of required switches and transceivers to achieve the desired connectivity. The total cost of each architecture is 
\begin{equation}
    \text{Total\ cost} = Price_{SW}\times N_{SW} + Price_{TR}\times N_{TR}
\end{equation}
while the power is
\begin{equation}
    \text{Total\ power} = Pwr_{SW}\times N_{SW} + Pwr_{TR}\times N_{TR}
\end{equation}

The last two columns of Table~\ref{tab:nw_swich} provide the cost and power savings of a Rail-only interconnect over the state-of-the-art. Our Rail-only design notably reduces the network cost by 38\% to 77\% (117 to 234 million dollars) and power consumption by 37\% to 75\% (1.7 to 6.9 megawatts) compared to the state-of-the-art design while achieving equivalent performance.
This reduction stems from eliminating spine layer switches and decreasing the number of switch tiers within each rail. Surprisingly, switches with a radix of 64 provide the worst-case cost and power reduction in both cluster sizes. In this case, the state-of-the-art design requires a three-tier \fattree network, while the Rail-only design requires two tiers for each rail. Still, our design only requires three-quarters of the total number of switches while achieving the same performance as the state-of-the-art design. 

\begin{table}[t]
\scriptsize
\centering
\caption{Number of switches and transceivers for different clusters.}

\renewcommand{\arraystretch}{1} 
\linespread{1.05}\selectfont\centering
    \begin{tabular}{|P{0.55cm}|P{0.55cm}|P{0.55cm}|P{0.65cm}|P{0.65cm}|P{0.65cm}|P{0.55cm}|P{0.55cm}|}
    \hline
    \multirow{2}{*}{\#GPUs} & \multirow{2}{*}{{\shortstack{Switch\\ Radix}}} & \multicolumn{2}{c|}{\#Switches ($N_{SW}$)} & \multicolumn{2}{c|}{\#Transceivers ($N_{TR}$)} & \multicolumn{2}{c|}{Savings} \\\cline{3-8}
       & & \multicolumn{1}{c|}{SOTA} & \multicolumn{1}{c|}{Rail-only} & \multicolumn{1}{c|}{SOTA} & \multicolumn{1}{c|}{Rail-only} & \multicolumn{1}{c|}{Cost} & \multicolumn{1}{c|}{Pwr} \\
    \hline
        \multirow{3}{*}{32768}  
                                 & \multicolumn{1}{c|}{64} & \multicolumn{1}{c|}{2560} & \multicolumn{1}{c|}{1536} & \multicolumn{1}{c|}{196608} & \multicolumn{1}{c|}{131072} & \multicolumn{1}{c|}{38\%}  & \multicolumn{1} {c|}{37\%} \\\cline{2-8}
                                 & \multicolumn{1}{c|}{128} & \multicolumn{1}{c|}{1280} & \multicolumn{1}{c|}{256} & \multicolumn{1}{c|}{196608} & \multicolumn{1}{c|}{65536} & \multicolumn{1}{c|}{77\%} & \multicolumn{1}{c|}{75\%} \\\cline{2-8}
                                 & \multicolumn{1}{c|}{256} & \multicolumn{1}{c|}{384} & \multicolumn{1}{c|}{128} & \multicolumn{1}{c|}{131072} & \multicolumn{1}{c|}{65536} & \multicolumn{1}{c|}{62\%} & \multicolumn{1}{c|}{60\%} \\\hline
        \multirow{3}{*}{65536}   
                                 & \multicolumn{1}{c|}{64} & \multicolumn{1}{c|}{5120} & \multicolumn{1}{c|}{3072} & \multicolumn{1}{c|}{393216} & \multicolumn{1}{c|}{262144} & \multicolumn{1}{c|}{38\%} & \multicolumn{1}{c|}{37\%} \\\cline{2-8}
                                 & \multicolumn{1}{c|}{128} & \multicolumn{1}{c|}{2560} & \multicolumn{1}{c|}{1536} & \multicolumn{1}{c|}{393216} & \multicolumn{1}{c|}{262144} & \multicolumn{1}{c|}{38\%}  & \multicolumn{1}{c|}{37\%} \\\cline{2-8}
                                 & \multicolumn{1}{c|}{256} & \multicolumn{1}{c|}{1280} & \multicolumn{1}{c|}{256} & \multicolumn{1}{c|}{393216} & \multicolumn{1}{c|}{131072} & \multicolumn{1}{c|}{77\%} & \multicolumn{1}{c|}{75\%} \\\hline
    \end{tabular}
    \label{tab:nw_swich}
\end{table}
\section{Related Work} \label{sec:discussions}

\para{LLM trend.}  
The current growth rate of LLM computational and speed requirement outpaces the advancements in AI accelerators and network speed as Moore's law slows down, necessitating hyper-scale clusters 
and more efficient interconnects~\cite{sirius, dnnmodelgrowth}. The MegatornLM line of work pioneers LLM parallelization~\cite{shoeybi2020megatronlm,narayanan2021efficient,korthikanti2022reducing}. Our position to remove any-to-any network connectivity complements MegatronLM.
We also acknowledge ongoing efforts to reduce language models' size and resource requirements without compromising performance~\cite{dolly}. These works complement our design as our design reduces network resources and maintains performance even for smaller language models and clusters. Similarly, research directions that aim to directly reduce the amount of communication through quantization and compression, like DeepSpeed Zero++, also complement our approach~\cite{deepspeedzeropp}.

\para{LLM Inference.} This paper explores the training workload of LLMs, yet inference represents another significant part of the LLM product cycle. Inference demands fewer resources as it involves moving less data through the LLM and only computes the forward pass and multiple passes to generate response tokens~\cite{alpaserve}. Pope et al. developed specific parallelism for inference on TPU-v4 architecture~\cite{pope2022efficiently}. For our design, each HB domain becomes an inference-serving domain with low latency, and the Rail-only connections help load-balance multiple inference domains. We leave a detailed investigation of LLM inference to future work. \looseness=-1 

\para{Multi-job training.} It is common for a GPU cluster to train multiple smaller jobs simultaneously. Existing works focus on \fattree-based GPU clusters and provide techniques for better fairness and shorter job-completion time~\cite{gandiva,tiresias,muri,rajasekaran2023cassini}. While this paper focuses on training a single LLM on a large cluster, the Rail-only network design is also suitable for a multi-job setting. The entire cluster can be arbitrarily partitioned by tiling into smaller rectangular partitions, similar to the case of TPU-v4~\cite{jouppi2023tpu}. Each partition then independently executes a smaller training job.  \looseness=-1 

\para{ML infrastructures and other ML workloads.} Prior works illustrated the benefits of co-designing hardware and software for ML models. For instance, Google's TPU cluster is optimized for training large models with 3D parallelism on TPUs~\cite{jouppi2023tpu}, while Meta's Neo focuses on training recommendation models with large embedding tables~\cite{mudigere2023softwarehardware}. Our work focuses on designing a cost-efficient network to train LLMs efficiently. Although our proposed Rail-only architecture focuses on network design specifically for LLMs, our design is efficient for many other DNN workloads when combined with other efforts, as the forwarding overhead is low~(\S\ref{sec:routing}). Recent works attempt to make the parallelization strategy and collective communication algorithms bandwidth-aware for any DNN model~\cite{alpa, unity, zhao2023bandwidth}, producing traffic patterns ideal for the Rail-only network. \looseness=-1

\vspace*{-1mm}
\section{Conclusion} \label{sec:conclusion}
In this paper, we examine and analyze the traffic pattern of LLM training with hybrid parallelism. We propose a novel \textit{Rail-only architecture} that aligns with LLMs' distinct characteristics and demands. Our architecture leads to 38\% to 77\% cost reductions and 37\% to 75\% power savings while maintaining identical performance to the state-of-the-art GPU networks.

\section*{Acknowledgments}
We thank anonymous Hot Interconnects reviewers for their feedback. The MIT-affiliated authors are supported by DARPA FastNICs 4202290027, NSF SHF-2107244, NSF CAREER-2144766, NSF PPoSS-2217099, NSF CNS-2211382, NSF FuSe-TG-2235466, Sloan fellowship FG-2022-18504, ACE and CUbiC centers sponsored by Semiconductor Research Corporation (SRC) and DARPA under the JUMP 2.0 program.

\bibliographystyle{IEEEtran}
\bibliography{ref}

\end{document}